\documentstyle[12pt,aaspp4]{article}

\newcommand{\etal}{{\it et al.}\ }
\newcommand{\eg}{{\it e.g.}\ }
\newcommand{\cf}{{\it cf.}\ }

\begin{document}

\title{On the origin of the Trojan asteroids:  Effects of Jupiter's mass 
accretion and radial migration}

\author{Heather J. Fleming and Douglas P. Hamilton}

\begin{center}

Astronomy Department, University of Maryland\\[40pt]

Heather J. Fleming\\
Astronomy Department\\
University of Maryland\\ 
College Park, MD 20742-2421\\
Tel: (301) 474-7582\\  
Fax: (301) 314-9067\\
E-mail: hcohen@astro.umd.edu\\[10pt]

Douglas P. Hamilton\\
Astronomy Department\\
University of Maryland\\ 
College Park, MD 20742-2421\\
Tel: (301) 405-1548\\  
Fax: (301) 314-9067\\
E-mail: hamilton@astro.umd.edu\\[60pt]

Submitted to {\it Icarus}: June 16, 1999\\
Revised: \today\\[60pt]

41 pages\\[7pt]
13 Figures and 2 Tables\\[40pt]

\end{center}
\newpage

\begin{center}
Running head:  Origin of the Trojan asteroids\\[40pt]

Corresponding Author:\\

Douglas P. Hamilton\\
Astronomy Department\\ 
University of Maryland\\
College Park, MD 20742-2421\\ 
Tel: (301) 405-1548\\  
Fax: (301)314-9067\\
E-mail: hamilton@astro.umd.edu\\ 
\end{center}

\newpage

\doublespace

%

\begin{abstract}

  We present analytic and numerical results which illustrate the
  effects of Jupiter's accretion of nebular gas and the planet's
  radial migration on its Trojan companions.  Initially, we
  approximate the system by the planar circular restricted three-body
  problem and assume small Trojan libration amplitudes.  Employing an
  adiabatic invariant calculation, we show that Jupiter's thirty-fold
  growth from a $10 M_\oplus$ core to its present mass causes the
  libration amplitudes of Trojan asteroids to shrink by a factor of
  about 2.5  to $\sim 40\%$ of their original size.  The calculation
  also shows that Jupiter's radial migration has comparatively little
  effect on the Trojans; inward migration from 6.2 to 5.2 AU causes an
  increase in Trojan libration amplitudes of $\sim4\%$.  In each case,
  the area enclosed by small tadpole orbits, if made dimensionless by
  using Jupiter's semimajor axis, is approximately conserved.  Similar
  adiabatic invariant calculations for inclined and eccentric Trojans
  show that Jupiter's mass growth leaves the asteroid's eccentricities
  and inclinations essentially unchanged, while one AU of inward
  migration causes an increase in both of these quantities by
  $\sim 4\%$.

  Numerical integrations confirm and extend these analytic results.
  We demonstrate that our predictions remain valid for Trojans with
  small libration amplitudes even when the asteroids have low, but
  nonzero, eccentricities and inclinations and/or Jupiter has an
  eccentricity similar to its present value.  The integrations also
  show that Trojans with large libration amplitudes, including
  horseshoe orbits, are even more strongly affected by Jupiter's mass
  growth and radial migration than simple scaling from our analytic
  results would suggest.  Further, the numerical runs demonstrate that
  Jupiter's predicted mass growth is sufficient to cause the capture
  of asteroids initially on horseshoe orbits into stable tadpole
  orbits.  Thus, if Jupiter captured most of its Trojan companions
  before or while it accreted gas, as seems probable, then Jupiter's
  growth played a significant role in stabilizing Trojan objects by
  systematically driving them to lower libration amplitudes.

\vspace{1.0in}

Celestial Mechanics\\
Jupiter\\ 
Origin, Solar System\\
Resonances\\

\end{abstract}
\newpage

\section{Introduction}

The Trojans are a distant group of asteroids dynamically linked to
Jupiter by a 1:1 mean motion resonance which causes them to librate
about stable Lagrangian equilibrium points located $60^\circ$ in front
of (L4) and behind (L5) Jupiter along its orbit.  Because Trojans
orbit far from the Sun and also have low albedos, they suffer from a
low discovery rate and hence are underrepresented among numbered
asteroids.  Extrapolating from the identified Trojan objects,
Shoemaker \etal (1989) estimate that the total Trojan population
contains nearly half as many asteroids as the main belt.  The large
number of Trojans and their strong dynamical connection to Jupiter
make determining their origins an important goal; by understanding the
early history of the Trojans, we may also gain insight into Jupiter's
formation and early evolution.

Clues about the origins of the Trojan asteroids may be found in their
current physical and orbital properties, which include the overlapping
signatures of mechanisms which captured them into librating orbits, as
well as processes which have contributed to the population's evolution
over time.  Some distinctive characteristics of the current Trojan
asteroids include a small mean eccentricity of $\sim 0.06$, a small
mean libration amplitude of $\sim 29^\circ$, and a large mean
inclination of $\sim 18^\circ$ (\cite{shoetal89}, \cite{levetal97}).
Also, nearly twice as many asteroids have been observed librating
about the L4 point as about the L5 point.  This, however, may simply
be the result of observational selection effects (\cite{shoetal89}).

There are many competing theories for the origin and evolution of the
Trojan asteroids.  It has been suggested that the Trojans may have
originally been comets (\cite{rabe72}) or near-Jupiter planetesimals
(\eg \cite{shoetal89}, \cite{kl95}).  A number of mechanisms have been
considered for capturing these objects into Trojan orbits, including
collisions between objects, drag forces, and mass accretion by
Jupiter.  Shoemaker \etal (1989) theorized that collisional
emplacement of fragments of near-Jupiter planetesimals during the
dispersion of the planetesimal swarm may have provided most of the
Trojan objects.  More recently, numerical modeling of the collisional
evolution of the Trojan population (Marzari \etal 1997) has shown that
collisions are largely responsible for shaping the current size
distribution of the smaller Trojans, as well as having caused the
escape of some of the Trojan objects into chaotic orbits.  Long-term
numerical integrations by Levison \etal (1997) have also shown that
the Trojan population is dynamically unstable and that dynamical
diffusion has contributed and continues to contribute to the loss of
Trojan objects.

Various types of drag forces acting on Jupiter and/or the Trojan
precursors have also been examined.  Kary and Lissauer (1995)
considered Solar nebular gas drag.  They showed numerically that gas
drag could cause planetesimals to be captured into 1:1 resonance with
a protoplanet.  Interestingly, they found that such capture is rare
for planets on circular orbits, but quite common for planets with
appreciable eccentricities.  Gas drag may also have played a
significant role in evolving the Trojan population into its present
form, provided that the Trojan precursors were captured before the
dispersion of the solar nebula (Peale 1993).  Yoder (1979) looked at
the effects of dynamical friction during Jupiter's dispersal of the
planetesimal swarm, which caused a slight inward migration of Jupiter.
He argued that this migration would cause a decrease in the libration
amplitudes of Jupiter's Trojan companions (see, however, Table~I).

\bigskip
\centerline{\fbox{\bf Insert Table I}}

The possibility that a change in Jupiter's mass could be responsible
for the capture of the Trojan asteroids was investigated by Rabe
(1954) who argued analytically that a decrease in Jupiter's mass could
cause its satellites to move onto Trojan orbits.  More recently,
Marzari and Scholl (1998) showed numerically that an increase in
Jupiter's mass could cause the capture of planetesimals into librating
orbits.  They used a proto-Jupiter on a ``best guess'' orbit growing
simultaneously with Saturn over a period of $10^4$ or $10^5$ years,
and found that a large fraction of the planetesimals initially on
horseshoe orbits and a small percentage of those initially orbiting
near the 1:1 resonance were captured into tadpole orbits.

Past changes in Jupiter's mass have also long been considered as a
potentially significant evolutionary mechanism for creating the
current distribution of Trojan asteroids; however, attempts to predict
the exact form of the effects of Jupiter's growth on the Trojans have
thus far been contradictory.  For an increase in Jupiter's mass,
Rabe (1954) theorized that the Trojan libration amplitudes would
increase, Horedt (1974a, 1974b, 1984) argued that they would not be
appreciably affected, and Yoder (1979) argued that they would decrease
(Table~I).

In this paper, we focus on the changes which Jupiter underwent early
in its history.  We investigate the significance of Jupiter's mass
growth and radial migration as Trojan capture mechanisms and
especially as mechanisms for evolving the Trojan population.  We focus
on these two processes in isolation in order to fully characterize
their behavior. Other potentially important processes including gas
drag, the gravity of other planets, and collisions amongst Trojans are
not considered here, because models which include all of these effects
would have a large number of poorly-determined free parameters. For
example, with gas drag, what is the gas density as a function of
distance from the Sun? When and exactly how does Jupiter form a gap in
the gas distribution? How sensitive are Trojan asteroids to different
gas drag models? These questions are important and need to be studied
in depth.  There are many open questions like these in the full
Trojan formation problem, probably more than can be addressed in a
single paper.  Accordingly, we have chosen to study individual
processes first for later incorporation into a more general model. A
strong advantage to this approach is that investigating individual
processes in detail will ultimately lead to a deeper physical
understanding of results from more complicated models.

Adopting this approach, we first present consistent analytic and
numerical results which show the effects of Jupiter's growth and
radial migration on its Trojan companions in the limit of a
slowly-changing Jupiter, thus settling the previous controversy. We
then explore the working of these mechanisms for a wide range of
timescales, initial Trojan libration amplitudes, Jupiter
eccentricities, and asteroid eccentricities and inclinations.

\section{Analytic Results}

\subsection{Libration Amplitude}

Consider the planar circular restricted three-body problem, in which
two massive bodies move about each other in circular orbits due to
their mutual gravitation and a third body of infinitesimal mass moves
in the orbital plane of the two massive objects.  This system admits
five equilibrium points where a test particle can have zero velocity
and zero acceleration in the frame which corotates with the primary
masses about their common center of mass (Danby 1988).  Three of these
points lie along the line through the two primaries and are unstable.
The other two, L4 and L5, lie at the tips of the equilateral triangles
whose bases are the line connecting the primary masses (see Fig.~1).
These are called the triangular Lagrangian equilibrium points and are
stable to small oscillations so long as the mass ratio of the
primaries, $\mu = {M_2 / (M_1 + M_2)}$ (where $M_1$ and $M_2$ are the
larger and smaller of the primary masses, respectively), satisfies
$\mu \lesssim 0.0385$ (\cite{md99}).  This condition is met for all
Sun-planet and planet-moon pairs in the Solar system, with the
exception of Pluto and Charon.

\bigskip
\centerline{\fbox{\bf Insert Figure 1}}

The planar circular restricted three-body problem is a reasonable
approximation for the system consisting of the Sun, Jupiter, and a
Trojan asteroid, since the asteroid's mass is insignificant in
comparison to either Jupiter or the Sun, and Jupiter's eccentricity is
relatively small ($\sim0.0483$ currently).  If we make the further
approximations that the Trojan's oscillations about its equilibrium
position are small and that the asteroid is on a nearly circular
orbit, then, noting that
\begin{equation}
\mu = {M_J \over {M_S + M_J}} \approx 0.001 \ll 1,
\end{equation}
\noindent
where $M_S$ is the mass of the Sun and $M_J$ is the mass of Jupiter,
the motion of the asteroid is well approximated by the equation
\begin{equation}
\ddot \phi + \left({ 27 \over 4 }\right) \mu n_J^2 \phi = 0
\label{Tmotion}
\end{equation}
\noindent
(\cite{bs64}), 
where $\phi$ is the difference between the mean longitudes of the asteroid and 
Jupiter, 
\begin{equation}
n_J = \left[{ {G(M_S + M_J)} \over {a_J^3} }\right]^{1/2}
\label{nJ}
\end{equation} 

\noindent
is the mean motion of Jupiter ($a_J$ is Jupiter's semimajor axis and
$G$ is the gravitational constant), and an overdot signifies
differentiation with respect to time.  The observed librational
motions of the Trojan asteroids are well approximated by the solution
of Eq.~\ref{Tmotion},

\begin{equation}
\phi = {A \over 2} \cos(\omega t + B),
\label{Tsolution}
\end{equation}
\noindent
where 
\begin{equation}
\omega^2 = {27 \over 4}\mu n_J^2,
\label{omega}
\end{equation}
\noindent  
$t$ is time, and $A$ and $B$ are constants.  Note that $A / 2$ is the
amplitude of the $\phi$ oscillations.  The libration amplitude, $A$, is
defined to be the total angular extent of these oscillations (see
Fig.~1).  Equivalently, the system can be described by the
Hamiltonian:
\begin{equation}
H = {1 \over 2} a_J^2 \dot \phi^2 + {1 \over 2} \omega^2 a_J^2
\phi^2 .
\label{Thamiltonian}
\end{equation}
\noindent
The canonical variables for this Hamiltonian are $q = a_J \phi$ and 
$p = a_J \dot \phi$, so that ${\partial H \over {\partial p}} = \dot q$, and
${\partial H \over {\partial q}} = - \dot p$
reproduces the equation of motion (Eq.~\ref{Tmotion}).

When changes are made to this system (\eg mass growth of one of the primaries
or the addition of an external drag force), the Hamiltonian is no longer 
conserved.
However, if the changes to the Hamiltonian system are slow enough, it can
be shown that the action, $J = \int p dq$, is approximately conserved 
(\cite{ll60}).  Such slow changes are called adiabatic changes, and $J$ is 
an adiabatic invariant.  Using the expressions
for $p$, $q$, and $\phi$ above, we determine the action for the three-body 
system:
\begin{equation} 
J_{3body} = \int a_J^2 \dot\phi d\phi = 
\int_0^{2\pi/\omega} a_J^2 A^2 \omega^2 \sin^2(\omega t + B) dt = 
{\pi \over 4} \sqrt{27 G \over 4} A^2 M_J^{1/2} a_J^{1/2} = constant
\label{Taction}
\end{equation}
\noindent
where we have evaluated the integral by using the adiabatic approximation 
that $a_J$, $A$, and $\omega$ are constant over one libration period.
The conservation of the action can be written in the useful form:
\begin{equation}
{A_f \over {A_i}} = \left({ M_{Jf} \over {M_{Ji}}} \right)^{-1/4}
\left({ a_{Jf} \over {a_{Ji}}}\right)^{-1/4}
\label{analyticresult}
\end{equation}
\noindent
where the subscripts $i$ and $f$ indicate the initial and final values of the
variables, respectively.

For radial migration of Jupiter, we can find the resulting change in
the Trojan's libration amplitude directly from
Eq.~\ref{analyticresult}; since changing $a_J$ has no effect on $M_J$,
the factor $(M_{Jf}/M_{Ji})^{-1/4}$ is equal to 1.  Then, for a
physically reasonable inward radial migration of Jupiter from
approximately 6.2 AU to 5.2 AU (see Section 4),
Eq.~\ref{analyticresult} predicts an increase in the Trojan's
libration amplitude of only $\sim4\%$.

Determining the effects of Jupiter's mass growth on the Trojan's
libration amplitude is a bit more subtle, since increasing $M_J$ may
affect $a_J$ as well as $A$.  In order to determine how changing $M_J$
affects $a_J$, we digress briefly to discuss a second adiabatic
invariant calculation. If we consider the two-body system of Jupiter
and the Sun moving about one another on elliptical orbits due to their
mutual gravitation, we can calculate an invariant action for slow
changes in the mass of either body.  Noting that, in this case,
adiabatic means that changes to the system are negligible over one
orbital period, we find

\begin{equation}
J_{2body} = 2\pi \sqrt{G (M_S + M_J) a_J} = constant .
\label{2Bconstant}
\end{equation}
\noindent
This result was contemplated nearly one hundred years ago when Str\"omgen 
(1903) considered what effect mass accretion by the Earth 
would have on the orbit of the Moon (see also \cite{jeans61}).  
For our case, the invariance of $J_{2body}$ means that if the mass of either 
Jupiter or the Sun were slowly decreased, Jupiter's orbit would drift outward.
This effect is well known (\eg Horedt 1984) and was observed recently in 
numerical simulations by Duncan and Lissauer (1998) in
which the orbits of the outer planets were seen to expand as the mass
of the Sun was decreased to a small fraction of its original value.
Similarly, if either Jupiter or the Sun slowly accretes
mass, Jupiter's semimajor axis will decrease.  Note that Eq.~\ref{2Bconstant}
implies that adding a Jovian
mass of material to either Jupiter or the Sun produces the same
change in the semimajor axis of the system.
  
Returning to our discussion of the effects of Jupiter's mass growth on
its Trojan companions, we see from Eq.~\ref{2Bconstant} that if we
change the mass of Jupiter adiabatically, the semimajor axis of
Jupiter's orbit will be altered according to
\begin{equation}
{a_{Jf} \over a_{Ji}} = { {M_S +M_{Ji}} \over {M_S + M_{Jf}}}.
\label{amj}
\end{equation}
\noindent
Substituting this into Eq.~\ref{analyticresult}, we see the full effect which
altering $M_J$ has on the Trojan's libration amplitude:
\begin{equation}
{A_f \over A_i} = \left({ {M_{Jf} \over M_{Ji}} }\right)^{-1/4} 
\left({ 
{M_S + M_{Jf}} \over {M_S + M_{Ji}}
}\right)^{1/4}.
\label{LibAmpmj}
\end{equation}
\noindent
Since $M_J \ll M_S$, the second term in parentheses is very nearly
equal to 1, and the change in $A$ is given sufficiently accurately by
Eq.~\ref{analyticresult} if we simply take $(a_{Jf}/a_{Ji})^{-1/4} =
1$.  We find that the growth of Jupiter by gas accretion from a $\sim
10 M_\oplus$ core to its present mass of $\sim 320 M_\oplus$ causes $A$
to decrease to $\sim40\%$ of its original value.  Thus, if librating
Trojan asteroids were already present when Jupiter was a $10 M_\oplus$
core, then the effects of Jupiter's mass increase would be substantial
and would dominate over the effects of its radial migration. The
effect of Jupiter's radial migration is comparable to its mass
accretion only if the planet moves inward by a factor of $\sim 30$,
\eg from 150 AU to 5 AU, substantially more than current theories
predict.

There has been some significant confusion in the literature about the
effects of Jupiter's mass growth on the libration amplitudes of its
Trojan companions (see Table~I).  Our analytic result for the effects
of Jovian mass growth disagrees with the findings of Rabe (1954) and
Horedt (1974a, 1974b, 1984), but agrees exactly, in both direction and
magnitude, with the conclusions of Yoder (1979).  Our prediction for
the effects of Jupiter's radial migration, however, disagrees with
Yoder's 1979 result (see Table~I).  Statements in Yoder \etal (1983)
about the tidal evolution of the Saturnian satellites Janus and
Epimetheus, however, are inconsistent with Yoder's 1979 calculations,
but agree, at least in sign, with our radial migration results.  In
order to dispel this confusion, we carefully verify our analytic
predictions with numerical simulations in sections 3 and 4 below.

We can make further use of the two-body adiabatic invariant for the 
Sun-Jupiter system (Eq.~\ref{2Bconstant}) to gain additional insight 
into the meaning of 
the three-body adiabatic invariant (Eq.~\ref{Taction}).  When $A$ and 
$\mu$ are both very small, the Trojan
asteroid orbit, viewed in the frame which corotates with Jupiter, looks
like a little ellipse centered on one of the Lagrangian equilibrium
points with the ratio of its semimajor to semiminor axis equal to
\begin{equation}
{ a \over b} = { 2 \over {\sqrt{3 \mu}}}
\label{ratio}
\end{equation}
\noindent
(Murray and Dermott 1999).  The area within this ellipse is 
\begin{equation}
area = \pi a b = {\pi \over 2} a^2 \sqrt{3 \mu} .
\label{area1}
\end{equation}
\noindent
For a small tadpole orbit, the libration amplitude times the semimajor axis
of Jupiter's orbit is approximately equal to 
the major axis of the ellipse ($A a_J \simeq 2a$).  Thus we can rewrite Eq.
\ref{area1} as
\begin{equation}
area = {\sqrt{3} \pi \over 8} \left({ M_J \over {M_S + M_J}}\right)^{1/2} A^2
a_J^2 = {2 \pi \over 3} {J_{3body} \over J_{2body}} a_J^2
\label{area2}
\end{equation}
\noindent
where we have used the two- and three-body adiabatic invariants 
defined in Eqs. \ref{2Bconstant} and \ref{Taction}, respectively.  So,
we find:
\begin{equation}
{area \over a_J^2} = constant.
\end{equation} 
\noindent
Thus, the invariance of the action for the three-body system implies
that the dimensionless area enclosed by a small-amplitude Trojan orbit
({\it i.e.,} the area enclosed by the Trojan orbit divided by the
square of Jupiter's semimajor axis), remains constant for an adiabatic
change to the system.  For example, as Jupiter accretes mass, the
resulting change in $a_J$ is relatively small, so as $A$ decreases,
the radial width of the Trojan orbit must increase in order to keep
the area nearly constant (see Fig.~1).  Note, however, that the area
within the tadpole orbit in Fig.~1 is not strictly conserved because
the libration amplitude of the tadpole is quite large.

Finally, we can use the two-body adiabatic invariant for the 
Sun-Jupiter system (Eq.~\ref{2Bconstant}) to determine what effect 
changing the mass of 
the Sun has on the libration amplitudes of Jupiter Trojans.  The dependence
of $A$ on $M_S$ is present in Eq.~\ref{analyticresult} through the
factor $(a_{Jf}/a_{Ji})^{-1/4}$.  We know from $J_{2body}$ 
(Eq.~\ref{2Bconstant})
that if the mass of the Sun is altered adiabatically, $a_J$ will change 
according to 
\begin{equation}
{a_{Jf} \over a_{Ji}} = {M_{Si} + M_J \over {M_{Sf} + M_J}}.
\label{ams}
\end{equation}

\noindent
Combining this with Eq.~\ref{analyticresult}, we find
\begin{equation}
{A_f \over A_i} = \left( { a_{Jf} \over {a_{Ji}} }\right)^{-1/4}
= \left({ 
{M_{Sf} + M_J} \over {M_{Si} + M_J}
}\right)^{1/4}
\label{LibAmpms}
\end{equation}
\noindent
for an adiabatic change in $M_S$.
Thus, slowly increasing the mass of the Sun increases the libration 
amplitude of the Trojan, which is the opposite of the effect on $A$
caused by adding mass to Jupiter.

\subsection{Trojan inclination and eccentricity}

Additional adiabatic invariant calculations determine the effects of
changing Jupiter's mass and semimajor axis on a Trojan asteroid's
eccentricity, $e_a$, and inclination, $i_a$. For the case when $i_a
\ne 0$, we approximate the orbit of the Trojan as an inclined circle.
Then, the motion of the asteroid in the $z$-direction can be well
represented by simple harmonic motion with a restoring force equal to
the sum of the $z$-components of the gravitational forces of the Sun
and Jupiter acting on the asteroid.  If we assume that the asteroid's
inclination and libration amplitude are small, the distances between
Jupiter, the asteroid, and the Sun are all approximately equal to
$a_J$.  Then, the restoring force per unit mass acting on the asteroid
is
\begin{equation}
f = {-G(M_S + M_J) \over {a_J^3}} z,
\end{equation}
\noindent
where $z$ is the distance of the Trojan asteroid above the plane of
Jupiter's orbit.  Notice that the vertical oscillation frequency is the
same as Jupiter's orbital frequency (Eq.~\ref{nJ}).  The Hamiltonian
for the vertical motion is
\begin{equation}
H = {1 \over 2} \dot z^2 + {1 \over 2} {G(M_S + M_J) \over {a_J^3}} z^2
\end{equation}
\noindent
which has canonical variables $q = z$ and $p = \dot z$.  As in 
section 2.1, for an adiabatic change to the system, the action is conserved.
Noting that
here adiabatic means that changes to the system are negligible 
over one orbital period, we find 
\begin{equation}
J_{incl} = \int \dot z dz = \pi i_a^2 \sqrt{ G(M_S + M_J)a_J} = constant,
\label{inclconst}
\end{equation}
\noindent
where we have used the relation $z_{max} = a_J \sin{i_a} \simeq i_a
a_J$ and integrated over a full orbital period.  From this result, we
see that if an external force adiabatically changes the semimajor axis
of Jupiter's orbit about the Sun, the inclination of the Trojan will
change according to the relation
\begin{equation}
{i_{a f} \over {i_{a i}}} = \left({ a_{J f} \over {a_{J i}}}\right)^{-1/4} .
\label{inclaresult}
\end{equation}

If instead the mass of Jupiter or the Sun is varied, 
Eq.~\ref{inclconst} tells us
that some combination of $i_a$ and $a_J$ must change to keep the action
constant.  We determine how the variation is shared between $i_a$ and
$a_J$ by using the invariant action for the two-body Sun-Jupiter
system, $J_{2body}$ (Eq.~\ref{2Bconstant}).  Dividing Eq.~\ref{inclconst}
by Eq.~\ref{2Bconstant}, we find
\begin{equation}
i_a = constant
\label{inclmresult}
\end{equation}
\noindent
for an adiabatic change in either $M_J$ or $M_S$.

The adiabatic calculation for an eccentric asteroid orbit parallels
that for an inclined orbit, however the radial oscillation frequency
is given by 

\begin{equation}
n_{radial} = \left[{ {G(M_S - {27\over4} M_J)} \over {a_J^3} }\right]^{1/2}
\label{Rfreq}
\end{equation}

\noindent
(Murray and Dermott 1999, page 94) rather than by Eq.~\ref{nJ}.
Accordingly, the action, evaluated for small eccentricities has the
form

\begin{equation}
J_{ecc} = \int \dot r dr = \pi e_a^2 \sqrt{ G\left( M_S - {27\over4}
M_J \right) a_J} = constant,
\label{Jecc}
\end{equation}

\noindent
and $e_a$ will respond to an adiabatic change in $a_J$ according to
the relation

\begin{equation}
{e_{a f} \over e_{a i}} = \left( { a_{J f} \over {a_{J i}} }\right)^{-1/4}.
\label{eccaresult}
\end{equation}

\noindent
This result is in agreement with analytic work by Gomes (1997) who
showed that $e_a$ would increase when $a_J$ was decreased. For
variations in $M_S$ and/or $M_J$, we combine Eqs. \ref{2Bconstant} and
\ref{Jecc},  and find

\begin{equation}
e_a \left(1-{31\over16}{M_J\over M_S}\right) = constant.
\label{eccmresult}
\end{equation}

\noindent
Thus an increase in Jupiter's mass should lead to a slight
increase in a Trojan asteroid's eccentricity.

In summary we find, using the invariance of the action for adiabatic
changes to the Sun-Jupiter and Sun-Jupiter-Trojan systems, that if the
semimajor axis of Jupiter is decreased, the asteroid's libration
amplitude (Eq.~\ref{analyticresult}), eccentricity
(Eq.~\ref{eccaresult}), and inclination (Eq.~\ref{inclaresult}) will
all increase.  We find further that if we increase Jupiter's mass,
$a_J$ will decrease (Eq.~\ref{amj}) and the Trojan's libration
amplitude will decrease (Eq.~\ref{LibAmpmj}), its eccentricity will
increase slightly (Eq.~\ref{eccmresult}), and its inclination will
remain unchanged (Eq.~\ref{inclmresult}).  Finally, if the Sun's mass
is increased, $a_J$ will be decreased (Eq.~\ref{ams}) and the asteroid
will be dragged inward with Jupiter, $A$ will increase by the same
amount it would were $a_J$ altered by an external force
(Eq.~\ref{LibAmpms}), $e_a$ will decrease slightly
(Eq.~\ref{eccmresult}) and $i_a$ will not change
(Eq.~\ref{inclmresult}). These results are also summarized in Table~II.

\bigskip
\centerline{\fbox{\bf Insert Table II}}

\section{Numerical Work: Jupiter's Mass Growth by Accretion}

In this section and the next, we confirm the analytic results above
and explore their range of validity by numerically integrating the
three-body system consisting of the Sun, Jupiter, and a massless
asteroid.  The full equations of motion, consisting of the
gravitational force of each body acting on the other two, are
integrated in inertial coordinates.  We use Bulrisch-Stoer and
Runge-Kutta methods with adaptive stepsize (\cite{numrec}), having
initially decided against faster symplectic methods (\cite{wh91})
because of the difficulty in handling close approaches with Jupiter.
Although fast symplectic methods for handling close approaches do
exist (\cite{duetal98}), the speed of our routines was adequate.  We
ran extensive tests on our code, including checking that the
Bulrisch-Stoer and Runge-Kutta integrators converged to the same
solutions, producing plots which matched specific Trojan asteroid
orbits illustrated in Murray and Dermott (1999, pages 97 and 98), and
verifying that the Jacobi constant was conserved to sufficient
accuracy for the circular restricted three-body problem with constant
Jupiter mass and semimajor axis.

We begin our numerical exploration by integrating the three-body
system as Jupiter grows from $\sim10 M_\oplus$ to its current mass.
We experimented with growing Jupiter both exponentially ($M_J = M_{J
  i} e^{\alpha t}$) and linearly ($M_J = M_{J i} + \beta t$), where
$M_{J i}$ is the initial mass of Jupiter and $\alpha$ and $\beta$ are
constants.  We find, as expected from section 2, that our results were
not significantly affected by the form of mass growth so long as the
growth was slow enough to be adiabatic; accordingly all results
presented below are for an exponentially-growing Jupiter.
Additionally, orbits librating around the Lagrangian equilibrium point
L4 behaved similarly to orbits around L5, as suggested by our analytic
analysis.  Thus, the results presented below apply to objects
librating about either of these points.

\subsection{Circular Coplanar Orbits}
\subsubsection{Dependence on Mass Growth Timescales}

Placing the Sun, Jupiter, and the asteroid all on 
initially circular coplanar orbits, we
carry out a set of integrations in which Jupiter grows on 
timescales ranging from $10^2$ to $10^5$ years.  We monitor the 
changes to the asteroid's libration amplitude and plot our results
in Fig.~2.

\bigskip
\centerline{\fbox{\bf Insert Figure 2}}

For long mass growth timescales ($\sim 10^4$ years), the numerical
asteroid orbits agree well with the analytic prediction.  Initially,
the numerically-determined points track the analytic prediction
precisely, but as Jupiter's mass grows the points begin to scatter
more. The increase in amplitude of the oscillation of the numerical
points about the analytically-predicted line is due to our method of
calculating the libration amplitude.  We use an analytic approximation
(Yoder 1983, Shoemaker {\it et al.} 1989) which makes the following
assumptions: 1) a planar three-body system with all the objects on
circular orbits, 2) $M_a \ll M_J \ll M_S$, and 3) $A$ is very small.
The Jupiter-Trojan system is reasonably approximated by these
assumptions; however, as $M_J/M_S$ grows, the error in the
approximation increases, thereby causing an increased scatter of the
calculated points.  Furthermore, the changing mass of Jupiter itself
leads to additional effects which are not accounted for in the simple
theory; these effects are larger for more rapid growth timescales.
Runs with slower growth timescales of $\sim 10^5$ years (not shown)
exhibit the same behavior as the $10^4$ year runs shown here, since
both of these slow growth rates represent changes to the system that
are well within the adiabatic limit where Eq.~\ref{analyticresult} is
valid.

For the faster Jupiter growth rates ($10^2$ and $10^3$ years), the
numerical curves deviate substantially from the analytic prediction of
Eq.~\ref{analyticresult} (Fig.~2).  In these cases, significant
changes in the mass of Jupiter occur on timescales comparable to the
libration period of the Trojan asteroid, $T = {2\pi \over \omega}$
(with $\omega$ given by Eq.~\ref{omega}), which is $\sim 900$ years
for $M_J =10 M_\oplus$ and $\sim 150$ years for Jupiter's current
mass.  Thus, the change in the Trojan orbit depends on its initial
conditions, {\it i.e.}, exactly where along the tadpole orbit the
asteroid starts.  The initial conditions for the $10^3$-year timescale
run shown in Fig.~2 are chosen so that the curve exhibits maximum
deviation from the analytic result.  Note that the oscillations in
this curve are not due to inaccuracies in the analytic approximation
used to calculate the libration amplitude; they are real effects due
the asteroid's librational motion and occur at the libration frequency.

The dependence of short timescale runs on initial conditions is
clearly illustrated by the curves for two extreme choices of asteroid
starting point for the $10^2$-year timescale plotted in Fig.~2.  The
curve which is everywhere above the analytic line corresponds to an
asteroid which was started at the point along the tadpole orbit
farthest away from Jupiter (maximum $\phi$, see Fig.~1).  The basic
characteristics of this curve can be understood using a simple
harmonic oscillator analogy: we treat the librational motion of the
asteroid as a one-dimensional oscillation in the $\phi$ direction, and
the rapid growth of Jupiter in $10^2$ years is approximated as
instantaneous growth.  Using this analogy, growing Jupiter when the
asteroid starts at the farthest point from Jupiter is equivalent to
increasing the restoring force of the harmonic oscillator instantly
when the oscillator is at its maximum extension.  This affects the
period of the oscillations, but leaves the amplitude unchanged.  For
this reason, the upper $10^2$-year curve in Fig.~2 is initially
horizontal, indicating no change in the libration amplitude.  Since
increasing Jupiter's mass over $10^2$ years is not truly an
instantaneous change, the libration amplitude of the Trojan asteroid
eventually decreases as it begins to move toward Jupiter.

The $10^2$-year timescale curve in Fig.~2 which is everywhere below
the analytic line corresponds to an asteroid started at the point on
the tadpole orbit farthest from the Sun (see Fig.~1) where $\vert
\dot{\phi} \vert$ is maximum.  Again using the harmonic oscillator
analogy, this is equivalent to increasing the restoring force when the
oscillator has its maximum velocity.  This causes the amplitude of the
oscillations to decrease.  Thus, we see an initially rapid decrease in
the Trojan's libration amplitude which slows as the asteroid moves
closer to Jupiter.

In the adiabatic limit, the effect of the mass change 
is averaged over all points along the asteroid's orbit.
Thus, the decrease in the asteroid's libration amplitude for adiabatic 
growth lies between the two extremes just discussed, as can be seen in 
Fig.~2.

\subsubsection{Dependence on Libration Amplitude}

Next, we explore the role of the asteroid's initial libration size 
in determining how its libration amplitude will change as Jupiter's
mass grows.  Recall that the analytic prediction (Eq.~\ref{analyticresult})
was derived with the assumption of small libration amplitude.  We study
a set of tadpoles with different initial libration amplitudes; in
each integration the timescale for Jupiter's
mass growth is $10^5$ years, well within the adiabatic limit.  We find
that for initial libration amplitudes $\lesssim 50^\circ$, our numerical 
results agree well with the analytic prediction, as expected (Fig.~3). 
The curve for the $A_i \simeq 50^\circ$ tadpole shows a
slight deviation from the analytic result; tadpoles with larger initial
amplitudes show even greater departures.  As in Fig.~2, the spread of 
the numerical points is due to the analytic method used to determine 
the libration amplitude.

\bigskip
\centerline{\fbox{\bf Insert Figure 3}}

The larger tadpole orbits shrink more rapidly than
Eq.~\ref{analyticresult} predicts.  Furthermore, the larger the
libration amplitude is, the faster it shrinks, as can be seen by
examining the initial slopes of the curves for the $110^\circ$,
$130^\circ$, and $150^\circ$ tadpoles in Fig.~3.  This can be
understood by considering the effective potential in the corotating
frame (Fig.~1).  The effective potential changes much more steeply at
the head of the tadpole near Jupiter than at the tail which lies
further away (Erdi 1997).  The larger the tadpole orbit is, the
further it extends away from Jupiter, and the less steep is the
potential in which the tail end of the orbit lies.  When the potential
is flatter, the location of the turning point of the orbit changes by
a greater amount for the same change in energy, and so the orbits
whose tails lie in the shallowest potential, {\it i.e.} those with the
largest libration amplitude, shrink the fastest (see Fig.~4).  Once
the initially large tadpole orbits become small enough, they shrink at
the rate predicted by Eq.~\ref{analyticresult}.  This is illustrated
in Fig.~3 where, by the end of the integrations, all the curves are
tending to the same slope as the analytic result.

\subsubsection{Horseshoe Orbits}

We also examine the effects of Jupiter's mass growth on very large
``horseshoe orbits'' whose librations encompass both the L4 and L5
equilibrium points.  Figure~4 shows the results of an integration in
which the asteroid has an initial libration amplitude of $\sim
330^\circ$.  The horseshoe orbit shrinks slowly until $\sim 1.3 \times
10^4$ years when it transitions to an L4 tadpole.  The tadpole then
continues to shrink.

\bigskip
\centerline{\fbox{\bf Insert Figure 4}}

Although horseshoe orbits are well outside the range of libration
amplitudes described by our adiabatic calculation, they too shrink
under the influence of a growing Jupiter.  The rate of decrease in
libration amplitude is, however, different for horseshoe orbits and
tadpole orbits.  Looking at the lower edge of the plot in Fig.  4, we
see that the turning point of the horseshoe orbit pulls away from
Jupiter at a nearly constant rate for exponential growth of Jupiter's
mass.  Furthermore, the corresponding turning point in the tadpole
orbit (the tip which lies nearest Jupiter) pulls away approximately
linearly as well, but with a shallower slope.  For a doubling of
Jupiter's mass, one tip of the horseshoe orbit recedes from Jupiter by
$\sim 4.5^\circ$, while for the same mass change, the near-Jupiter tip
of the tadpole moves by only $\sim 3.0^\circ$.  In this way, 
horseshoe orbits shrink even faster than tadpole orbits.

Further integrations demonstrate that the growth of Jupiter to its
current mass captures most asteroids which were initially on horseshoe
orbits into tadpole orbits.  Figure~5 shows the results of a set of
integrations which determine the fate of asteroids placed into
different-sized horseshoe orbits at various points during the growth
of Jupiter.  We find that if the asteroids are placed on horseshoes
when Jupiter is a $10 M_\oplus$ core, orbits with initial libration
amplitudes as large as $ A \sim 346^\circ$ are captured into tadpole
orbits by the time Jupiter has reached its present mass. When the
asteroids are started on horseshoes after Jupiter has already grown
part-way to its current mass, fewer of the orbits shrink sufficiently
to transition to tadpoles; however, even if the asteroids are started
when Jupiter has reached three-quarters of its final
mass, some of the smaller horseshoe orbits still become tadpoles, at
least temporarily, during the course of the integration.

\bigskip
\centerline{\fbox{\bf Insert Figure 5}}

In Fig.~5, we can clearly see the chaotic nature of the evolution of
the horseshoe orbits: there is mixing between the orbits which escape
the system and those which remain in horseshoes for the entire time.
We also observe mixing at the boundary for capture into tadpoles.
Small differences in the initial asteroid orbits can significantly
alter the effects of Jovian perturbations, vastly changing the final
asteroid orbit.  Additionally, note that the smallest possible
horseshoe orbits occur at $A \sim 312^\circ$, nearly independent of
the mass of Jupiter.  This result is in agreement with the analytic
work of Horedt (1984).  The transition to a tadpole orbit occurs
whenever a horseshoe orbit shrinks to this minimum size.

Data from the integrations plotted in Fig.~5 confirms that the
observation made from Fig.~4, that each tip of a horseshoe orbit
recedes from Jupiter at a roughly uniform rate of $\sim 4.5^\circ$ per
Jupiter doubling, holds for general horseshoe orbits.  Figure~6 shows
the decrease in the libration amplitudes of the horseshoe orbits
plotted against the number of doublings of Jupiter's mass.  The solid
line, fit by eye to the data, has a slope of $9.0^\circ$ per doubling
(indicating that each of the two turning points recedes at $4.5^\circ$
per doubling).  The slight flattening of the data points near (0,0) on
the plot indicates that the horseshoe orbits shrink a bit more slowly
when they are near the transition to tadpole orbits.  We verified this
observation with additional simulations not shown here.  In order to
get the best value for the slope in Fig.~6, we ignored the points in
the flattened tail of the plot and did not insist that our fitted line
go through (0,0).

\bigskip
\centerline{\fbox{\bf Insert Figure 6}}

Next, we look at the final sizes of the tadpoles produced by the
transitions from horseshoes shown in Fig.~5 to see if these objects
could, in fact, contribute to the current population of Jupiter
Trojans.  In Fig.~7, we see that a small number of the asteroids
placed in horseshoe orbits when Jupiter is $\lesssim 10\%$ of its
final mass become tadpole orbits which are stable for a significant
fraction of the age of the Solar System.  A substantially larger
number of the horseshoe orbits started when Jupiter is $\lesssim 20\%$
of its final mass become tadpoles that remain stable for at least a
hundred million years.  Thus, assuming that objects resided on
horseshoe orbits when Jupiter was accreting mass, a small fraction of
these objects may still survive in the Trojan swarm today.
Furthermore, since these asteroids undergo relatively frequent
collisions (\cite{maretal97}), they almost certainly produced some
fragments which were ejected into more stable tadpole orbits.

\bigskip
\centerline{\fbox{\bf Insert Figure 7}}

\subsection{Eccentric and Inclined Trojan Orbits}

To further explore the range of validity of our analytic results from 
section 2 above, we add eccentricity and inclination to the Trojan orbit, at
first leaving Jupiter on a circular orbit.  We integrate the three-body system
for a number of different small values of eccentricity and inclination
($e_a \lesssim 0.1$, $i_a \lesssim 1^\circ$) with the Trojan on a small 
($A \sim 10^\circ$) tadpole orbit and Jupiter growing over $10^5$ years.  We 
find
that for these low values of eccentricity and inclination, our analytic
expression for the change in libration amplitude (Eq.~\ref{analyticresult})
still approximates the behavior of the system well.  Figure~8 shows
the results of one of our numerical integrations.  The
asteroid is given an initial eccentricity of $e_a \sim 0.01$ and an initial 
inclination of $i_a \sim 1^\circ$.  The analytic method of Yoder \etal (1983) 
for determining the libration amplitude, $A$, fails for non-zero $e_a$ and 
$i_a$, 
so we resort to a simpler but more time consuming method.
The values of libration amplitude are determined from the numerical data by 
taking the average of the local maximum and minimum values of $\phi$ at the
turning points ($\dot \phi = 0$) of the tadpole orbit.  The range in values 
of $\phi$ at the
turning points is due to both the eccentricity and inclination of the orbit.
The averaging process takes advantage of the difference in orbital and 
librational timescales, and is essentially the guiding
centre approximation for the orbit (Murray and Dermott 1999).  
The remaining uncertainty in the numerical
points is due primarily to the difficulty in determining the maximum and
minimum $\phi$ values from a sample of discrete
points.  Further, as the libration period
becomes smaller, we sample only parts of the elliptical orbit during
the turning point.  Despite these difficulties, we find that the numerical 
points in Fig.~8 follow the analytic curve well.

\bigskip
\centerline{\fbox{\bf Insert Figure 8}}

We also explore the effects of Jupiter's mass growth on the
eccentricity and inclination of Trojan orbits.  Figure~9 shows the
eccentricity and inclination of the Trojan orbit whose libration
amplitude is plotted in Fig.~8.  We see that, to first order, the
eccentricity and inclination are constant during Jupiter's growth, as
predicted in section 2.2.  A closer inspection reveals a slight
increase in the mean eccentricity which is in good agreement with the
analytical predictions of Eq. \ref{eccmresult}.

\bigskip
\centerline{\fbox{\bf Insert Figure 9}}

\subsection{Eccentric Jupiter Orbit}

Finally, we add eccentricity to Jupiter's orbit while keeping the Trojan
on an inclined and eccentric orbit.  Figure~10a shows the eccentricity and 
inclination of a Trojan when Jupiter has a constant mass and an
eccentricity, $e_J \simeq 0.05$.  The 
asteroid's eccentricity oscillates around a constant ``forced'' component 
which is equal to the eccentricity of Jupiter.  The ``free'' component of
the eccentricity is approximately the amplitude of the oscillations 
($\sim 0.05$ in Fig.~10a).  The Trojan's forced inclination is
zero, since the inclination is measured relative to 
Jupiter's orbit.  The high-frequency oscillations visible in the 
plot of $i_a$ are due primarily to the Trojan's librations, while the 
lower-frequency oscillations
in $i_a$ are correlated to the oscillations of the 
Trojan's eccentricity.

\bigskip
\centerline{\fbox{\bf Insert Figures 10a and 10b}}

Starting with the same initial orbits for both Jupiter and the Trojan, but
allowing Jupiter to accrete material over $2 \times 10^5$ years, we obtain the 
results
shown in Fig.~10b.  The values of the forced and free eccentricities and
inclinations are not significantly altered by Jupiter's growth; however,
the frequencies of the oscillations in both $e_a$ and $i_a$
increase as the mass of Jupiter grows, and the strength of its 
gravitational perturbation increases.  The slight drift in the mean 
Trojan inclination seen in Fig.~10b is simply part of a periodic
oscillation in $i_a$ correlated to the drift of the orbital node.  So, 
as predicted in section 2.2, we find that the eccentricity and 
inclination of a Trojan orbit remain essentially unchanged as 
Jupiter's mass grows, even when Jupiter is on an eccentric orbit.  Thus,
the primordial eccentricities and inclinations of some 
large Trojan objects may have been preserved during Jupiter's accretion
of mass.

\section{Numerical Work: Radial Migration of Jupiter}

We now turn to a series of integrations of the three-body
Sun-Jupiter-asteroid system in which Jupiter undergoes 1 AU of inward
radial migration. For our integrations, we set Jupiter's initial
semimajor axis to $\sim 6.2$ AU and leave its mass constant at $\sim
10 M_\oplus$.  Because the effects of Jupiter's radial migration are
independent of Jupiter's mass and depend only on the ratio of the
initial and final semimajor axes (see Eqs.  \ref{analyticresult},
\ref{inclaresult}, and \ref{eccaresult}) our numerical results are
applicable to migration of Jupiter at any point during its history.

The amount of radial migration that Jupiter underwent as a rocky core
and growing gas giant due to tidal interactions with the gas and
planetesimal disks is poorly constrained, but might be several AU
(\cite{wa97}).  Furthermore, after attaining its present mass, Jupiter
continued to experience radial migration due to scattering of
planetesimals by the giant planets (dynamical friction); recent models
for the formation of the Oort Cloud (\cite{fi96}, \cite{hm99}) predict
several tenths of an AU of radial migration at this stage. Our results
may be scaled to either or both of these scenarios.

We artificially cause Jupiter's orbit to shrink by applying a drag
force of the form ${\bf F} = - k {\bf v_J}$ (where ${\bf v_J}$ is
Jupiter's heliocentric velocity and $k$ is the drag constant) which
acts only on Jupiter.  This form of drag affects Jupiter's semimajor
axis, but not its eccentricity, providing a ``simplest case'' for
studying the effects of Jupiter's radial migration on Trojan objects.
Recall that our analytic work shows that, in the adiabatic limit,
changes in the Trojans' libration amplitudes are independent of the
exact form of the drag force (see Eq. \ref{analyticresult}); thus our
choice for the form of the drag force is reasonable.  As in the
integrations for Jupiter's mass growth, we find no apparent dependence
on the choice of Lagrangian equilibrium point, so the results below
are equally applicable to orbits about the L4 and L5 points.

\subsection{Circular Coplanar Orbits}
\subsubsection{Dependence on Radial Migration Timescale}

We initially place the Sun, Jupiter, and the asteroid all on circular
coplanar orbits.  We carry out a set of integrations with the Trojan
on a small tadpole orbit using different drag coefficients, $k$, to
cause Jupiter to migrate inward by $\sim 1$ AU on timescales ranging
from $\sim 10^2$ to $\sim 10^5$ years.  For slow evolution, we observe
that the Trojans are always dragged inward with Jupiter.  As in the
case of Jupiter's mass growth (section 3.1.1), we find that for
timescales significantly larger than the Trojan libration period
($\sim 1000$ years for $M_J = 10 M_\oplus$ and $a_J = 6.2$~AU), the
libration amplitude of the Trojan increases in the manner predicted by
Eq.~\ref{analyticresult}.  Also, as in the mass accretion case, when
the migration timescale approaches the libration period, the change in
libration amplitude deviates from our analytic prediction, with the
direction and amount of deviation depending on the asteroid's initial
librational phase.

For very fast migration rates, we find that the
asteroid is not pulled inward with Jupiter, but instead is ejected from 
its tadpole orbit.  For migration of Jupiter by $\sim 1$~AU in $10^3$~years, 
one sample integration shows an initially small tadpole orbit
transforming into a horseshoe orbit near the end of the
integration.  In another example, we change Jupiter's semimajor axis
in 500 years and find that the asteroid escapes entirely from the 1:1 
resonance after only $\sim 300$ years.

\subsubsection{Dependence on Libration Amplitude}

Next, we do a set of integrations starting the Trojan asteroid on
different-sized tadpole orbits.  In each run, we cause Jupiter to move
from $\sim 6.2$ to $\sim 5.2$~AU over $10^5$ years so that its
migration is adiabatic.  The results of several of these integrations
are shown in Fig.~11.  It is difficult to produce a plot similar to
Fig.~3 because Yoder's (1979) formula for calculating the libration
period assumes that there are no drag forces.  Thus, we estimate the
initial and final libration amplitudes by measuring the difference
between the minimum and maximum values of $\phi$ for the first and
last complete librations in the integration.  We note that the Trojans
librate about shifted equilibrium points due to the effects of the
drag force (\cf Murray 1994).  In Fig.~11 we see that for small tadpoles
($A_i \lesssim 30^\circ$) the numerical results agree well with the
prediction of Eq.~\ref{analyticresult}. The steplike appearance of the
first five points reflects difficulties inherent in our measurement
technique. As the initial libration amplitude becomes larger, however,
the numerical points deviate increasingly from the analytic result.
This is expected since the larger tadpoles break the assumption of
small libration amplitude which was made during the derivation of
Eq.~\ref{analyticresult}.  As in the case of Jupiter's mass growth, we
see that for larger initial tadpoles, the change in libration
amplitude is greater in magnitude but in the same direction as is
predicted analytically.  Also, we see that the deviation of $A_f/A_i$
from the analytic prediction increases more steeply as the initial
orbits become even larger.  As was discussed in section 3.1.2, this is
caused by the shallow slope of the effective potential near the tail
end (the end farthest from Jupiter) of large tadpole orbits.

\bigskip
\centerline{\fbox{\bf Insert Figure 11}}

\subsection{Eccentric and Inclined Trojan Orbits}

Next, we add eccentricity and inclination to the asteroid orbit,
leaving Jupiter on a circular orbit.  As in the case of Jupiter's mass
growth, we find that small Trojan eccentricities and inclinations ($ e
\lesssim 0.1$, $i \lesssim 1^\circ$) do not cause the behavior of the
libration amplitude to deviate significantly from the analytic
prediction of Eq.  \ref{analyticresult}.  However, unlike the mass
growth case, Jupiter's migration does affect the eccentricity and
inclination of the Trojan.  Figure~12 shows the semimajor axis,
eccentricity, and inclination of a Trojan asteroid on a small tadpole
orbit as Jupiter migrates from $\sim 6.2$ to $\sim 5.2$ AU in $10^5$
years.  During that time, the asteroid's inclination increases from
$1.000^\circ$ to $1.045^\circ$ and its eccentricity grows from 0.00999
to 0.01043, giving us $i_{a f}/i_{a i} = 1.045 \pm 0.001$ and $e_{a
  f}/e_{a i} = 1.044 \pm 0.001$.  These agree well with the analytic
prediction that $i_{a f}/i_{a i} = e_{a f}/e_{a i} = 1.045$ which we
calculate using Eqs. \ref{inclaresult} and \ref{eccaresult}.  These
changes in the asteroid's eccentricity and inclination resulting from
$\sim 1$ AU migration of Jupiter are quite small, but
are systematic.

\bigskip
\centerline{\fbox{\bf Insert Figure 12}}

\subsection{Eccentric Jupiter Orbit}

Finally, we explore the general case of non-zero Jovian eccentricity
and a Trojan on an eccentric and inclined orbit.  Figure~13 shows the
semimajor axis, eccentricity, and inclination of a sample Trojan
asteroid as Jupiter, with $e_J \sim 0.05$, migrates from $\sim 6.2$ to
$\sim 2.6$~AU in $3 \times 10^5$ years. Note that we allow Jupiter to
move by a much greater amount than in our earlier simulations so that
the effects of migration can be more easily observed and quantified.
As in section 3.3, the asteroid has a forced eccentricity equal to
Jupiter's eccentricity and a forced inclination of zero.  The asteroid
in Fig.~13 also has a free component of its eccentricity which is
initially $\simeq 0.004$, and a free inclination which is initially
$\simeq 1^\circ$.

\bigskip
\centerline{\fbox{\bf Insert Figure 13}}

As Jupiter migrates inward, the free components of both the asteroid's
eccentricity and inclination increase systematically.  From Fig.~13,
we find $i_{(free) a f}/i_{(free) a i} = 1.243 \pm 0.001$ and
$e_{(free) a f}/e_{(free) a i} = 1.32 \pm 0.15$.  Both of these
results agree well with the analytic prediction that $i_{a f}/i_{a i}
= e_{a f}/e_{a i} = 1.243$ (Eqs. \ref{inclaresult} and
\ref{eccaresult}) for an adiabatic change in Jupiter's semimajor axis
from 6.2 to 2.6 AU.  Thus, our analytic results (Eqs.
\ref{inclaresult} and \ref{eccaresult}) hold even when Jupiter has a
low eccentricity if $e_a$ and $i_a$ are interpreted as the free
components of the asteroid's eccentricity and inclination.

\section{Discussion}

We have shown with both a simple adiabatic calculation and numerical
simulations, that slow changes to the mass and semimajor
axis of Jupiter cause the Trojan libration amplitude to vary according 
to the relation:
\[
{A_f \over {A_i}} = \left({ M_{Jf} \over {M_{Ji}}} \right)^{-1/4}
\left({ a_{Jf} \over {a_{Ji}}}\right)^{-1/4}
\]
\noindent
for Trojans with small libration amplitudes, small eccentricities, and
small inclinations, and Jupiter with a small eccentricity.  For
inclined and eccentric Trojan objects, we find that Jupiter's mass
growth does not significantly affect either the Trojan's eccentricity
or inclination; however, Jupiter's radial migration causes a change in
the free component of both of these quantities by a factor of $\left({
    a_{J f} \over {a_{J i}}}\right)^{-1/4}$.

Applying our results to the core accretion model for the early
evolution of Jupiter, we find that the planet's growth by gas
accretion from a $\sim 10 M_\oplus$ core to its present mass would
cause a decrease in the libration amplitude of any Trojan companions
on small tadpole orbits to $\sim 40\%$ of their original size.  Our
representative choice for Jupiter's radial migration from $\sim 6.2$
to $\sim 5.2$~AU would result in an increase in the Trojans' libration
amplitudes, eccentricities, and inclinations of only $\sim 4\%$.  Even
for radial migrations of several tens of AU, the effects of Jupiter's
mass growth dominate over the effects of its migration.  Thus the
combined result of mass accretion and radial migration is to stabilize
Trojan objects by systematically driving them to lower libration
amplitudes.  Also, our numerical integrations show that the libration
amplitudes of Trojans on larger orbits shrink at an even faster rate.
Further, the shrinking of horseshoe orbits due to Jupiter's growth
could place additional objects or perhaps fragments of objects onto
stable tadpole orbits.  Thus, Jupiter's growth by mass accretion most
likely played a significant role in the capture and evolution of the
Trojan asteroid population.

Our results for the evolution of Trojan libration amplitudes,
eccentricities, and inclinations are quite general and can be applied
to other objects within the Solar System.  For example, Eqs.
\ref{analyticresult}, \ref{inclaresult}, and \ref{eccaresult} predict
that $A$, $e$, and $i$ will all decrease substantially if the
secondary body undergoes significant outward radial migration.  Uranus
and Neptune probably underwent more substantial radial migration due
to dynamical friction with planetesimals than Jupiter, moving outward
by as much as several AU (\cite{fi96}, \cite{hm99}).  This would have
caused a decrease in $A$, $e_a$, and $i_a$ of possible Trojan-like
companions by about $10\%$.  Radial migration effects might be even
more significant for some planetary satellites, notably our Moon.  The
Moon is believed to have formed via a giant collision which produced a
temporary ring of debris around the Earth (\cite{ce95} and
\cite{ietal97}).  Such a process would have most likely captured some
debris in librating orbits about the Moon's Lagrangian equilibrium
points.  Over the subsequent $4.5 \times 10^9$ years, the Moon
migrated outward to about 30 times its initial orbital radius.
Ignoring other effects, this migration should have decreased the
libration amplitudes, eccentricities, and inclinations of the debris
particles to $\sim 40\%$ of their original values, stabilizing these
objects in 1:1 resonance with the Moon.  Since we observe no such
objects today, either the 1:1 resonance was never populated, or other
effects (such as Solar gravity) caused them to be unstable.

Another potential area for study is the satellite system of Saturn.
The many resonances in this system are believed to have formed during
the significant outward migration of these satellites due to both
tidal interactions with Saturn and ring torques.  One unexplained
characteristic of the current Saturnian system is that of the six
largest satellites near Saturn, the middle two, Tethys and Dione, have
a total of three Trojan companions but the others, Mimas, Enceladus,
Rhea, and Titan, have none.  This is curious because there is no
obvious reason why the middle satellites should be the best at
capturing Trojan companions.  Also, the results of this paper suggest
that the objects which migrate outward by the greatest amount should
be the best at capturing and stabilizing their Trojan companions.
This suggests that the inner two satellites, which have migrated
farthest, should be most likely to have Trojan companions.  It is
likely that the probability of Trojan capture and the stabilization of
Trojan orbits is complicated by the presence of resonances between the
satellites.  The inner four Saturnian satellites are all currently
locked in resonances with each other, and they may have passed through
various other resonances in the past.  An exploration of the
interactions between the migration process explored in this paper, and
the effects of other resonances, discussed by Morais and Murray
(1999), may provide insight into this unexplained characteristic of
the Saturnian system.  It may also provide insight into how the
unusual pair of coorbital satellites, Janus and Epimetheus, which
librate on horseshoe orbits, was formed in the Saturnian system.

Our work is also directly relevant to the capture of planetary
satellites during the growth of the giant planets, as was suggested by
Heppenheimer and Porco (1977).  Equation~\ref{2Bconstant} clearly
shows that distant satellites would be drawn inward as a planet grows.
Their migration would cease when the planet reached its final mass.
Thus, this mechanism provides a natural way for a giant planet to pull
in distant satellites without causing them to collide with the planet.

\newpage

\singlespace

\begin{center}
Table I: Effects of Jovian Accretion and Migration on Trojan Libration
Amplitudes


\begin{tabular}{c||c|c|}

& if planetary mass increased 
& if inward radial migration \\
\hline\hline

Rabe (1954) & $A$ increases & X \\
\hline
Horedt (1974a, 1974b, 1984) & $A$ unaffected & X \\

\hline

Yoder (1979) & $A$ decreases & $A$ decreases \\

\hline

this work & $A$ decreases & $A$ increases \\

\hline

\end{tabular}

\vspace{.5in}
\end{center}


\begin{center}
Table II: Reaction of Orbital Parameters to \\
Imposed Adiabatic Changes in $a_J$, $M_J$, and $M_S$


\begin{tabular}{c||c|c|c|}

& if $a_J$ slowly decreased   & if $M_J$ slowly increased  & if $M_S$ slowly increased\\
\hline
\hline

then $a_J$ & X & decreases (Eq.~\ref{amj}) & decreases (Eq.~\ref{ams}) \\ 
\hline

then $A$ 
& increases (Eq.~\ref{analyticresult}) 
& decreases (Eq.~\ref{LibAmpmj})
& increases (Eq.~\ref{LibAmpms})\\

\hline

then $e_a$ 
& increases (Eq.~\ref{eccaresult})
& slightly increases (Eq.~\ref{eccmresult}) 
& slightly decreases (Eq.~\ref{eccmresult})\\

\hline

then $i_a$ 
& increases (Eq.~\ref{inclaresult}) 
& is unaffected (Eq.~\ref{inclmresult})
& is unaffected (Eq.~\ref{inclmresult})\\

\hline
\end{tabular}

\vspace{.5in}

\end{center}

\newpage

\doublespace
\normalsize

\begin{center}
{\bf Figure Captions}
\end{center}

\begin{center}
{\bf Fig.~1}
\end{center}
\vspace{-.2in}

Three sample Trojan asteroid orbits are plotted in the frame which
corotates with Jupiter about the center of mass of the Sun-Jupiter
system.  The L4 (leading) and L5 (trailing) Lagrangian equilibrium
points are each indicated by the symbol $\times$.  The orbits shown,
which enclose either the L4 or L5 point, but not both, are called
tadpoles due to the shape of their librations.  The tadpole around the
L4 point was integrated with Jupiter at twice its present mass, which
makes it wider and thus easier to view.  The libration amplitude, $A
\simeq 40^\circ$, is indicated for this orbit.  The two tadpoles about
the L5 point are the initial (long, thin tadpole with $A \simeq
85^\circ$) and final (short, fat tadpole with $A \simeq 60^\circ$)
orbits for a Trojan as Jupiter's mass grows slowly from one-half to
twice its current value.  The tadpole both shortens and widens as
Jupiter's mass increases.

\begin{center}
{\bf Fig.~2}
\end{center}
\vspace{-.2in}

The Trojan libration amplitude, $A$, (normalized to its initial value,
$A_i$) is plotted against Jupiter's mass, $M_J$, (normalized to its
initial value, $M_{Ji}$) on log-log scale for a series of integrations
during which Jupiter grows from $\sim 10 M_\oplus$ to its current
$\sim 320 M_\oplus$.  The initial Trojan orbits are small ($A \simeq
10^\circ$) tadpoles.  The analytic prediction of
Eq.~\ref{analyticresult} is plotted as a solid line.  The numerical
curve for the $10^4$ year growth rate (dotted line) agrees well with
the analytic result.  Curves for longer growth timescales (not shown)
agree equally well. The curves for shorter timescales, however,
deviate significantly from the analytic curve in a manner which
depends on the initial conditions of the orbit.  Note that the two
curves for the $10^2$-year timescale differ only in the initial
librational phase of the orbit.

\newpage

\begin{center}
{\bf Fig.~3}
\end{center}
\vspace{-.2in}

The Trojan libration amplitude (normalized to its initial value) is
plotted on a log-log scale for a variety of different-sized initial
tadpole orbits as Jupiter grows from $\sim 10 M_\oplus$ to its present
mass.  Our analytic result (Eq.~\ref{analyticresult}) is plotted as a
heavy solid line.  The upper dotted curve, which overlays the
theoretical curve, is for a tadpole with $A \simeq 50^\circ$
initially.  All smaller tadpoles (not shown) agree even better with
the analytic prediction.  The lower curves in this figure are tadpoles
with (from top to bottom) $110^\circ$, $130^\circ$, and $150^\circ$
initial libration amplitudes.  The large tadpole orbits shrink faster
than our analytic work predicts.  The oscillations in the numerical
curves are a result of the analytic approximation used to calculate
the libration amplitude and occur at the libration frequency.

\begin{center}
{\bf Fig.~4}
\end{center}
\vspace{-.2in}

Here we show the behavior of $\phi$, the longitude of the asteroid in
the frame which corotates with Jupiter, as Jupiter grows from $\sim 10
M_\oplus$ to its present mass in $3 \times 10^4$ years.  Jupiter is at
$\phi = 0^\circ$, and the initial asteroid orbit is an $A \sim
330^\circ$ horseshoe.  At $\sim 1.3 \times 10^4$ years the orbit jumps to
an L4 tadpole with $0^\circ \le \phi \le 180^\circ$.  The Trojan's
libration period is initially about $\sim 1,200$ years, and decreases
to $\sim 600$ years when the horseshoe transfers to a tadpole.
Afterwards, the tadpole's libration period continuously shortens in
accordance with Eq.~\ref{omega}.  Note that the tadpole orbit shrinks
most rapidly when its tail ({\it i.e.} its away-from-Jupiter turning
point) is near $\phi = 180^\circ$, since the effective potential is
flattest there.  Further, observe that the lower edge of the plot has
a steeper slope for the horseshoe orbit than for the tadpole orbit;
the straight line overlaying the plot is fit by eye to the edge of the
horseshoe orbit and has a slope of $\sim 4.5^\circ$ per Jupiter mass
doubling.

\begin{center}
{\bf Fig.~5}
\end{center}
\vspace{-.2in}

This plot shows the final state of asteroids started on horseshoe
orbits at different points during Jupiter's growth to its current size
of $320 M_\oplus$.  The vertical axis shows the initial libration
amplitude of the horseshoe orbit and the horizontal axis shows the
mass of Jupiter when the asteroid was placed in that orbit.  Asteroids
which escaped from the 1:1 resonance during the $10^5$ years of the
integration are indicated by open circles.  Objects which remained in
horseshoe orbits for the entire integration are shown as filled
triangles, and asteroids which were captured into tadpole orbits are
shown as stars. Most, but not all, of the orbits which
became tadpoles remained tadpoles for the rest of the integration.
The mixing of final states shown in this figure is an indication of
the chaotic nature of the orbits.

\begin{center}
{\bf Fig.~6}
\end{center}
\vspace{-.2in}

This plot shows the decrease in the libration amplitudes of those
horseshoe orbits from Fig.~5 which did not escape the 1:1 resonance
plotted against the number of secondary mass doublings.  A line with a
slope of $9.0^\circ$ per doubling was fitted by eye to the data.  The
low scatter of the points about this line indicates that the decrease
in $A$ by $9.0^\circ$ per doubling is a general property of horseshoe
orbits, independent of the mass of the secondary.

\begin{center}
{\bf Fig.~7}
\end{center}
\vspace{-.2in}

This plot shows the final libration amplitudes of the asteroids from
Fig.~5 which were on tadpole orbits at the end of the $10^5$-year
integration.  The horizontal axis is the mass of Jupiter at the time
when the asteroids were placed into their initial horseshoe orbits.
The orbits represented by points below the long-dashed line are stable
for more than $10^9$ years, while those below the short-dashed line
are stable for greater than $10^8$ years.  These timescales are
estimated from the work of Levison \etal (1997) for tadpoles with zero
eccentricity.

\newpage

\begin{center}
{\bf Fig.~8}
\end{center}
\vspace{-.2in}

The square data points show our numerical measurements of the change
in the libration amplitude of a Trojan asteroid with $A_i \simeq
10^\circ$, $e_a \simeq 0.01$, and $i_a \simeq 1^\circ$ as Jupiter (on
a circular orbit) grows from $\sim 10 M_\oplus$ to its current mass in
$10^5$ years.  Representative error bars are shown for two data
points.  The solid line is the analytic prediction for $e_a = i_a = 0$
(Eq.~\ref{analyticresult}).  The numerical points agree with the
analytic curve to within the error bars.

\begin{center}
{\bf Fig.~9}
\end{center}
\vspace{-.2in} 

This plot shows the eccentricity and inclination of the tadpole orbit
whose libration amplitude is plotted in Fig.~8.  The mean values of
the asteroid's eccentricity and inclination are essentially unchanged
(note the vertical scale) by the growth of Jupiter's mass, although
the tiny increase in the mean eccentricity is real. The mean
eccentricity rises by approximately one small tick mark over $10^5$
years, which is consistent with the $\sim0.2$\% increase predicted by
Eq.~\ref{eccmresult}.  The high frequency oscillations visible in the 
traces of both $e_a$ and $i_a$ are due to the Trojan's librational 
motion and thus become more rapid as Jupiter's mass increases, in 
accordance with Eq.~\ref{omega}.  The low frequency oscillations in 
the inclination are correlated to the precession of the asteroid's 
pericenter (not shown) driven by the disturbing effects of Jupiter. 
Since all the oscillations are caused by Jovian perturbations, they 
increase in amplitude as Jupiter's mass grows. 

\begin{center}
{\bf Fig.~10}
\end{center}
\vspace{-.2in}

These plots compare the orbital evolution of a Trojan asteroid a)
without and b) with Jovian mass growth.  For both cases, the Trojan
asteroids start with identical initial conditions and Jupiter has an
eccentricity of $\sim 0.05$. In a), the mass of Jupiter is kept
constant at $\sim 10 M_\oplus$, while in b) Jupiter's mass grows from
$\sim 10 M_\oplus$ to its current mass in $2 \times 10^5$ years.  Both
$e_a$ plots show oscillations about a forced eccentricity equal to
Jupiter's eccentricity.  The plots of $i_a$ show the free inclination
(since $i_{(forced) a} = 0$).  The low frequency oscillations in $i_a$
are correlated to the oscillations in $e_a$.  The high frequency
oscillations visible in the plots of $i_a$ are due to the Trojan's
librations.  The free and forced eccentricities and inclinations of
the Trojan asteroid are essentially unchanged by Jupiter's growth, as
predicted in section 2.2; however, the frequencies of all the observed
oscillations in $e_a$ and $i_a$ increase as Jupiter's mass grows.

\begin{center}
{\bf Fig 11}
\end{center}
\vspace{-.2in}

This plot shows the ratio of final to initial libration amplitude for
several different-sized tadpole orbits affected by the migration of
Jupiter from $\sim 6.2$ to $\sim 5.2$~AU over $10^5$ years.  Jupiter
is on a circular orbit.  The results of numerical integrations are
shown as solid dots with error bars which reflect the difficulties in
measuring the libration amplitude.  The horizontal line at $A_f/A_i =
1.045$ represents the analytic prediction of Eq.~\ref{analyticresult}
which is valid only for small initial libration amplitudes.  The
numerical results agree well with the analytic prediction for $A_i
\lesssim 30^\circ$, but deviate increasingly from the prediction for
larger initial libration amplitudes.

\begin{center}
{\bf Fig.~12}
\end{center}
\vspace{-.2in}

This plot shows the change in the semimajor axis, eccentricity, and
inclination of a Trojan asteroid on a slightly eccentric and inclined
orbit as Jupiter (on a circular orbit) migrates radially from $\sim
6.2$ to $\sim 5.2$~AU in $10^5$ years.  The Trojan's eccentricity and
inclination increase by a factor of $(a_{J f}/a_{J i})^{-1/4} =
(6.2/5.2)^{-1/4} = 1.045$ as is predicted by Eqs. \ref{inclaresult}
and \ref{eccaresult}.  Note that the tiny oscillations in $e_a$ and
$i_a$ are due primarily to the Trojan's librational motion.

\begin{center}
{\bf Fig.~13}
\end{center}
\vspace{-.2in}

This plot shows the semimajor axis, eccentricity, and inclination of a
Trojan asteroid on a slightly eccentric and inclined orbit when
Jupiter has an eccentricity of $\sim 0.05$ and migrates radially from
$\sim 6.2$ to $\sim 2.6$~AU in $3 \times 10^5$ years.  The forced
components of both the asteroid's eccentricity and inclination remain
constant ($e_{(forced) a} \sim 0.05$ and $i_{(forced) a} = 0$);
however, the free components of both $e_a$ and $i_a$ increase like
$(a_{J f}/a_{J i})^{-1/4}$, as predicted by Eqs. \ref{inclaresult} and
\ref{eccaresult}.

\end{document}